\documentclass[10pt]{article}

\usepackage{amsmath}
\usepackage{amssymb}
\usepackage{bm}
\usepackage{graphicx}

\usepackage{cite}

\usepackage{color} 

\usepackage{setspace} 

\usepackage[labelfont=bf,labelsep=period,justification=raggedright]{caption}

\makeatletter
\renewcommand{\@biblabel}[1]{\quad#1.}
\makeatother

\date{}

\begin{document}

\begin{flushleft}
{\Large
\textbf{An integrative approach for modeling and simulation of Heterocyst pattern formation in Cyanobacteria strands}
}
\\
Alejandro Torres-S\'anchez$^{1,2}$, 
Jes\'us G\'omez-Garde{\~n}es$^{1,3}$, 
Fernando Falo$^{1,3,\ast}$
\\
\bf{1} Departamento de F\'{\i}sica de la Materia Condensada, Universidad de Zaragoza, E-50009 Zaragoza, Spain.
\\
\bf{2} Laboratori de C\`alcul Num\`eric, Universitat de Polit\`ecnica de Catalunya, E-08034 Barcelona, Spain.
\\
\bf{3} Instituto de Biocomputaci\'on y F\'{\i}sica de Sistemas Complejos (BIFI), Universidad de Zaragoza, E-50018 Zaragoza, Spain.
\\
$\ast$ E-mail: fff@unizar.es
\end{flushleft}

\section*{Abstract}

A comprehensive approach to cellular differentiation in cyanobacteria is developed. To this aim, the process of heterocyst cell formation is studied under a systems biology point of view. By relying on statistical physics techniques, we translate the essential ingredients and mechanisms of the genetic circuit into a set of differential equations that describes the continuous time evolution of combined nitrogen, PatS, HetR and NtcA concentrations. The detailed analysis of these equations gives insight into the single cell dynamics. On the other hand, the inclusion of diffusion and noisy conditions allows simulating the formation of heterocysts patterns in cyanobacteria strains.
The time evolution of relevant component concentrations are calculated allowing for a comparison with experiments. Finally, we discuss the validity and the possible improvements of the model. 

\section*{Author Summary}
Understanding the underlying mechanisms favoring the cooperative behavior exhibited by different cell types constitutes the first step to explain the emergence of specialized cells in the early life stages of multicellular systems. Multicellularity appeared on Earth some billion years ago, and cyanobacteria are one of the first organisms that developed this feature. In fact, being differentiation processes the cornerstone of multicellularity, cyanobacteria strands are paradigmatic examples of prokaryotic cellular differentiation and cooperative pattern formation. When a strand of cyanobacteria cells is deprived of combined nitrogen, some of the vegetative cells start differentiating into heterocysts, which are  terminally differentiated nitrogen-fixing cells. However, not all  vegetative cells differentiate as (\emph{i}) heterocysts loose their photosynthetic capacity so they need vegetative cells around to be provided of a source of fixed carbon and (\emph{ii}) cell division, i.e.~reproduction, is only accomplished by vegetative cells. From such paradigmatic example it is clear that differentiation processes are the result of the interplay of complex regulatory networks acting inside the cell and external stimuli, from both the neighboring cells and the environment. Thus, in this article we present an integrative approach that combines the study of the internal regulatory processes, diffusion and noisy environments in order to capture the key mechanisms leading to the differentiation of vegetative cyanobacteria into heterocysts and the subsequent pattern formation.

\newpage
  
\section*{Introduction}

One of the most striking and complex problems tackled by biology is the formation of multicellular organisms. Multicellular organisms rely on cell differentiation, a mechanism by which cells become more specialized. More precisely, differentiation processes lead to precise alterations in gene expression that result in changes in cell morphology and function. These processes are highly dynamical, directed by complex regulatory networks involving cell-to-cell interactions, and often triggered by external stimuli. As a result of the differentiation, the organism separates functions in different cell types establishing a rich cooperative pattern that increases its complexity and adaptability. Due to the large number of scales involved, ranging from protein binding to diffusion of specific elements throughout the organism, a correct mathematical modeling of differentiation processes and biological pattern formation demands an integrated approach combining tools from statistical mechanics and the theory of dynamical systems (see for instance \cite{Koch1994,Suel2006}). \smallskip

A landmark process of (prokaryotic) cellular differentiation and cooperative pattern formation is the heterocyst differentiation in cyanobacteria strands \cite{wolk,flores}. Cyanobacteria are thought to be one of the first organisms in developing multicellularity some ($2-3$) billion years ago \cite{Schirrmeister2011}. These bacteria perform oxygenic photosynthesis releasing oxygen to the environment. On the other hand, nitrogenase, the enzyme that performs nitrogen fixation, is deactivated by oxygen so that nitrogen fixation cannot occur in its presence \cite{shi}. Cyanobacteria solve the incompatibility of incorporating both oxygenic photosynthesis and nitrogen fixation by separating these processes (\emph {i}) temporally, such as in the unicellular {\em Cyanothece sp.} strain ATCC 51142 that presents photosynthetic activity during the day and fixes nitrogen during the night \cite{toepel}, or (\emph{ii}) spatially, by the generation of non-photosynthetic nitrogen-fixing cells.
\smallskip

When provided of combined nitrogen (such as nitrate, nitrite, ammonium or urea), most cyanobacteria (with {\em Anabaena} PCC strain 7120 the most representative example) form long filaments of photosynthetic vegetative cells. However, in the absence of combined nitrogen (cN), a part of the vegetative cells differentiate into heterocysts, which are terminally differentiated nitrogen-fixing cells. By differentiating, heterocysts lose their photosysthetic capacity, so they need for an external source of fixed carbon \cite{Yoon1998,Zhang2006}. To this aim, each forming heterocyst sends a signal, by means of the diffusion of some particular chemical along the chain, to its neighboring cells to avoid their differentiation. A cooperative pattern is thus established: heterocysts provide cN to the cellular chain while non-differentiated cells remain supplying fixed carbon to the system. As a result, heterocysts appear interspersed within around 10 vegetative cells, depending on the species, forming a semiregular pattern that remains approximately constant in the chain regardless cell division \cite{flores}. This pattern forms one of the simplest and most primitive examples of multicellular organism as a product of the interdependence between heterocysts and vegetative cells. Interestingly, an isolated cyanobacterium does not differentiate but it first divides so that one of the descendants differentiates. This is necessary since (\emph{i}) a sole heterocyst would lack of a source of fixed carbon and (\emph{ii}) it would not reproduce since it is a terminally differenciated cell \cite{kumar}. 

\smallskip
Some quantitative modeling has been done to explain the dynamical and equilibrium properties of heterocyst pattern formation. In references \cite{allard,brown} Rutenberg and coworkers analyzed a model to explain heterocyst patterns by means of the study of combined nitrogen diffusion along a cyanobacterial strain. On the other hand, Gerdtzen {\em et al.} \cite{gerdtzen} modeled cyanobacterial filaments based on a time-discrete dynamical system exhibiting the main interactions between the most important proteins that take part in heterocyst formation. 

\smallskip
In this work, we develop a simple mathematical model, by integrating the recent experimental results on the gene regulatory network of cyanobacteria into the theoretical machinery of system biology. To this aim, we take into account the fundamental genetic paths that underlie the differentiation of heterocysts considering the interactions that come into play in this process. Furthermore, our model connects the diffusion of combined nitrogen with the dynamical properties of the underlying genetic circuit of each cyanobacterium, linking pattern formation and maintenance.  Our model shows that noise enables the development of the characteristic heterocysts patterns for a wide range of model parameters, revealing, in a quantitative way, that cyanobacteria strains may have evolved towards an efficient response mechanism to the noisy conditions that characterize the natural environment. 


\smallskip
The work is structured as follows.  First we  present the main actors of the basic regulatory network and the different dynamical interactions that take place during the differentiation process. Then we develop a mathematical model for the unicellular reaction to nitrogen deprivation. Although a unicellular model does not suffice to understand heterocyst formation, we analyze the main features that arise from the dynamical behavior of the system to gain insight about cell dynamics under different external conditions. 
Finally, we round off the paper by introducing the spatial model consisting of a filament of cyanobacteria, each one characterized by the dynamical circuit developed previously, that interact by means of protein diffusion.

\section*{Results}


\begin{figure}[t!]
\centering
\includegraphics[width=2.20in]{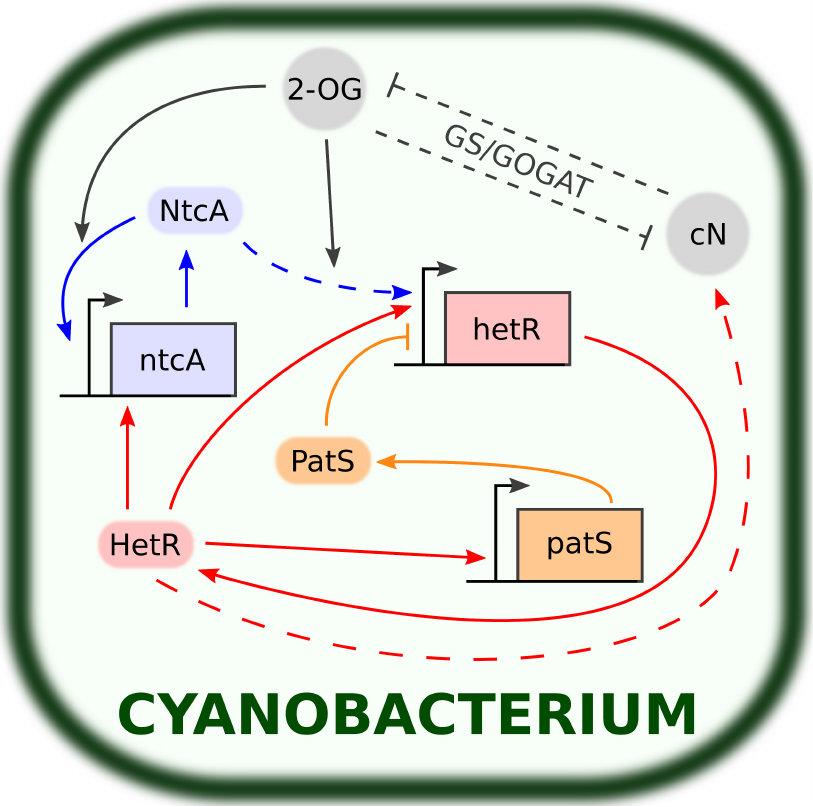}
\caption{Main components and interactions involved in the reaction to nitrogen deprivation in cyanobacteria. Rectangular boxes represent genes ({\it ntcA}, {\it hetR} and {\it patS}) while rounded boxes and circles represent transcription factors (NtcA, HetR and PatS) and smaller molecules (2-OG and cN) respectively. Normal-tipped and flat-tipped arrows stand for up-regulating and down-regulating processes respectively. Dashed lines stand for indirect or imperfectly understood interactions. 
The accumulation of 2-OG enhances the DNA-binding activity of NtcA, which in turn up-regulates the transcription of {\it ntcA} and {\it hetR}. HetR  activates {\it ntcA} and {\it hetR} (composing the central NtcA-HetR autoregulatory loop), the inhibitor {\it patS} and other genes that lead to nitrogen fixation and the morphological changes involved in heterocyst differentiation. 2-OG and cN levels are linked through the GS/GOGAT cycle (see Fig.~\ref{fig:gsgogat}).
}
\label{fig:model}
\end{figure}

\subsection*{Description of the main Genes and their basic Genetic Circuit}


Heterocyst development begins with sensing combined-nitrogen (cN) limitation and ends with nitrogen fixation in mature heterocysts. This process is usually completed after a time of about $20$ hours at 30$^\circ$C \cite{Zhang2006}. In Fig.~\ref{fig:model} we show a basic scheme of the genetic circuit including the most relevant elements and their respective interactions. Here we explain the main features of this genetic circuit.

 \smallskip

\begin{figure}[t!]
\centering
\includegraphics[width=2.75in]{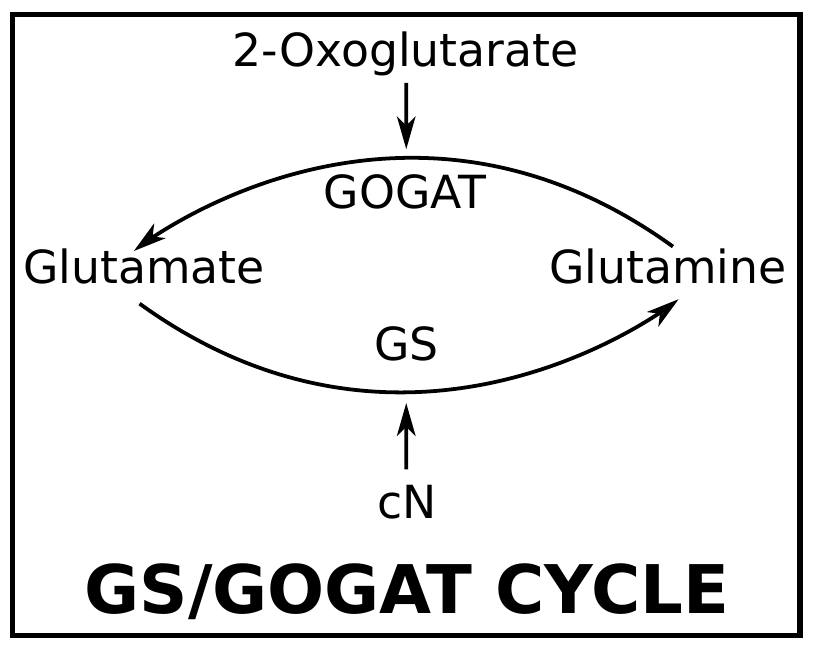}
\caption{2-OG and cN indirectly interact through the GS/GOGAT cycle. Glutamine is transformed into glutamate by means of 2-OG through the 2-OG amidotransferase (GOGAT) while cN converts glutamate into glutamine through the glutamine synthetase (GS). The importance of the cycle in heterocyst differentiation is twofold. From one side, it constitutes the early one-cell sensor to nitrogen starvation: the absence of cN breaks the cycle down and 2-OG starts to accumulate, whose action leads to the cascade of processes that provoke the differentiation (see Fig.~\ref{fig:model}). On the other hand, later during the differentiation, it process the cN created by the heterocysts decreasing the levels of 2-OG. The latter is crucial for the formation of the heterocyst pattern (see Fig.~\ref{fig:diffusion}).}
\label{fig:gsgogat} 
\end{figure}

The process is initiated with the  accumulation of  the enzyme 2-oxoglutarate (2-OG) as a consequence of nitrogen deprivation \cite{Laurent2005,Zhang2006}. 2-OG interacts with ammonium through the GS/GOGAT cycle \cite{Muro-Pastor2001,Vazquez-Bermudez2003,Muro-Pastor2005} (see Fig.~\ref{fig:gsgogat}). Under cN starvation, the GS/GOGAT cycle breaks down, leading to the afore-mentioned accumulation of 2-OG inside the cell \cite{Zhang2006}. In its turn, 2-OG stimulates the DNA-binding activity of NtcA, an important transcription factor for heterocyst development \cite{Vazquez-Bermudez2003, Wei1994, Frias1994}. Furthermore, the transcription of the genes targeted by NtcA does not start in the absence of 2-OG \cite{Vazquez-Bermudez2002,Tanigawa2002}. 
NtcA presents an auto-regulatory activity \cite{Ramasubramanian1996,Vazquez-Bermudez2002,Ramasubramanian1994} and indirectly activates the key gene that controls cell differentiation and pattern formation: \emph{hetR} \cite{Buikema1991,Buikema1991a,Ehira2006}. For its binding activity NtcA needs for homodimerization \cite{Wisen1999,Alfonso2001}. Therefore, the accumulation of 2-OG is the factor that triggers differentiation. In agreement with this idea, artificial increased levels of 2-OG result in heterocyst development even in the presence of ammonium\cite{Li2003,Vazquez-Bermudez2003,Laurent2005}.

 \smallskip

As already mentioned, the next step of the genetic circuit is the activation of \emph{hetR}. Importantly, null mutants of \emph{hetR} do not produce heterocysts whereas an overexpression of \emph{hetR} leads to an increased heterocyst frequency\cite{Buikema2001,Khudyakov2004,Buikema1991a}. The transcription of \emph{hetR} is induced by NtcA through the action of an intermediate, {\it nrrA} \cite{Ehira2006}. The DNA-binding activity of HetR requires its homodimerization \cite{Zhou1998,Huang2004}. Multiple transcription factors related to heterocyst formation are up-regulated by HetR, including {\it hetR} itself \cite{Huang2004}, {\it ntcA} \cite{Muro-Pastor2002} and {\it patS} \cite{Huang2004}. 

 \smallskip

\begin{figure}[t!]
\centering
\includegraphics[width=3.30in]{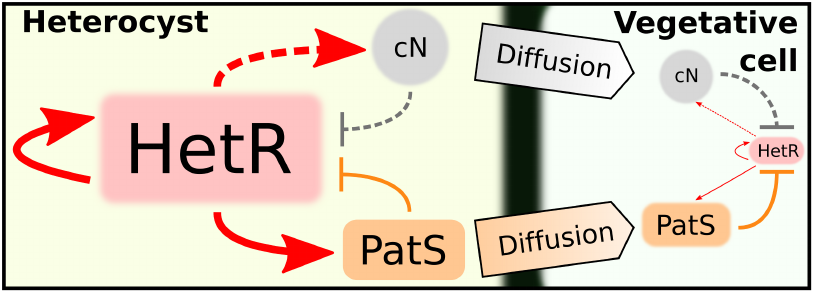}
\caption{Schematic representation of the diffusion processes that sustain the heterocyst pattern. Heterocysts produce cN and PatS. cN diffuses along the strain where, due to the action of the GS/GOGAT cycle (see Fig.~\ref{fig:gsgogat}), decreases the levels of 2-OG breaking the autoregulatory core NtcA-HetR. On the other hand, early during the differentiation, PatS (or other derivative of it, see the text) diffuses along the strain inhibiting HetR. Both processes combined prevent the differentiation of the rest of vegetative cells and explain the formation of the pattern.}
\label{fig:diffusion} 
\end{figure}

The up-regulatory loop composed of NtcA and HetR is central for heterocyst differentiation \cite{Muro-Pastor2002,Valladares2004,Fiedler2001,Hebbar2000}.  However,
the action of NtcA and HetR alone cannot account for pattern formation. Another transcription factor, PatS, inhibits the DNA-binding activity of HetR \cite{Huang2004,Wu2004,Yoon1998,Yoon2001}. This inhibitory behavior is essential for  the communication with adjacent cells and thus for achieving the observed patterns of vegetative cells and heterocysts in cyanobacteria strains (see Fig.~\ref{fig:diffusion}). Furthermore, {\it patS} is strongly expressed in differentiating cells and mature heterocysts due to its upregulation by HetR \cite{Yoon1998}. A strain without {\it patS} develops multiple contiguous heterocysts (about a $30\%$ of all cells as compared to the usual $10\%$ in the wild-type strain). On the other hand, an overexpression of {\it patS} suppresses heterocyst differentiation \cite{Zhang2006}. In fact, the addition to the growth medium of a synthetic peptide composed of the last five residues (RGSGR) of PatS (PatS5) inhibits heterocyst development, suggesting that PatS5 may be a diffusive mature form of PatS that stops the differentiation of the rest of vegetative cells of the strain \cite{Wu2004}.

 \smallskip

The last stages of heterocyst development cause the physiological changes of the cell aimed at creating an anaerobic environment that sustains nitrogen fixation. To this end, two new membrane layers are biosynthesized to decrease the entry of oxygen into the cell \cite{Fay1992}. The morphogenesis of these two layers is controlled by two family of genes, \emph{hep} and \emph{hgl}, that are indirectly up-regulated by HetR  \cite{Huang2004}. After these morphological changes the genes in charge of nitrogen fixation, {\it nif} genes, are expressed. These genes encode, among others, the nitrogenase enzyme that performs nitrogen fixation. 


\medskip

The fixed nitrogen of the new heterocysts, when transferred to other cells of the strain, acts as an inhibitor of their differentiation together with the transferred PatS \cite{Wolk1974}. Thus, the diffusion of both PatS and cN from heterocysts  along the strain plays a key role in its pattern maintenance (see Fig.~\ref{fig:diffusion}). Finally, after the differentiation process is finished, a cooperative system between heterocysts and vegetative cells is established so to ensure the survival of the strain. In particular, heterocysts produce fixed nitrogen from N$_2$ of the atmosphere and they interchange this nitrogen with the oxygen derivatives produced by the vegetative cells. 


\subsection*{Regulatory equations}

In this section we translate the genetic circuit described previously into a set of differential equations, for which we follow the derivation in \cite{hwa, Bintu2005, Phillips2012}. Details are left to supplementary information (SI). To simplify notation, constants related to NtcA, HetR, PatS and cN are denoted with the letters $a$, $r$, $s$, and $n$ respectively.

\medskip

We start looking at the transcription of \emph{ntcA}, which is regulated by HetR and NtcA itself. We assume that the probability that NtcA binds the promoter in the absence of 2-OG can be neglected. Taking into account that both HetR and NtcA dimerize to bind DNA we find:
\begin{equation}
v_a=L_a + \frac{v^a_a \kappa^a_a [\text{2-OG}][\text{NtcA}]^2+v_a^r \kappa_a^r [\text{HetR}]^2+v_a^{ar} \kappa_a^a\kappa_a^r[\text{2-OG}][\text{NtcA}]^2[\text{HetR}]^2}{ (1+\kappa_a^a [\text{2-OG}][\text{NtcA}]^2)(1+\kappa_a^r [\text{HetR}]^2)},
\label{ntca}
\end{equation}
where $v_a$ measures the production rate of NtcA in units of concentration per time, $v^a_a$, $v^a_r$ and $v^a_{ar}$ are the rates when only NtcA, only HetR or both are bound to DNA respectively, and $\kappa^a_*$ are the inverse of the effective dissociation constants of the compounds that bind DNA. On the other hand, $L_a$, the so-called leak term, measures the basal production of {\it ntcA} in the absence of regulation. Subscripts and superscripts identify the binding site and the transcription factor for which the constants are given respectively.
\medskip

Similarly we can obtain the transcription velocity for HetR. We assume that {\it hetR} is regulated by NtcA by means of a usual Hill function although the real process presents an intermediate, {\it nrrA}. This can be done if the mediator element(s) are not regulated through another factor of the genetic circuit under consideration and if they relax rapidly to their limiting values. On the other hand, PatS affects the autoregulatory loop of HetR. It has been suggested that PatS binds the binding site of HetR in the promoter of {\it hetR} preventing that HetR binds it \cite{Huang2004}. These facts, along with the influence of the levels 2-OG, provide with an expression for the transcription velocity:
\begin{equation}
v_r= L_r+\frac{v_r^a \kappa_r^a [\text{2-OG}][\text{NtcA}]^2(1+\kappa_r^s[\text{PatS}])+v_r^r \kappa_r^r [\text{HetR}]^2+v_r^{ar} \kappa_r^a\kappa_r^r[\text{2-OG}][\text{NtcA}]^2[\text{HetR}]^2}{ \left(1+\kappa_r^a [\text{2-OG}][\text{NtcA}]^2\right)\left(1+\kappa_r^r [\text{HetR}]^2+\kappa_r^s[\text{PatS}]\right)}
\label{hetr}
\end{equation}
HetR regulates most processes of the genetic circuit. It governs, among others, the transcription of {\it ntcA}, {\it patS},  {\it hep}, {\it hgl} and {\it nif} genes that lead to most of structural changes of the cell and to nitrogen fixation.
\medskip

The inhibitor PatS is regulated by HetR, and we assume no other influence. This gives the simple transcription velocity:
\begin{equation}
v_s= L_s+\frac{v_s^r \kappa_s^r [\text{HetR}]^2}{1+\kappa_s^r [\text{HetR}]^2}
\label{pats}
\end{equation}

Finally, we have to relate nitrogenase concentration [Ni] to that of combined Nitrogen [cN], both regulated by HetR and the levels of 2-OG [2-OG]. Let us begin by examining nitrogenase concentration, which is directly controlled by {\it nif} genes. Although this is not a direct process, we can assume, as we did for the NtcA-regulation by {\it hetR}, that {\it nif} genes are functionally governed by [HetR] following a typical Hill function. The nitrogenase production rate is given by:
\begin{equation}
        \frac{d \text{[Ni]}}{dt}=L_{Ni}+\frac{v_{Ni}^r \kappa_{Ni}^r \text{[HetR]}^2}{1+ \kappa_{Ni}^r \text{[HetR]}^2}-\delta_{Ni} \text{[Ni]}.
        \label{mo5}
\end{equation}
where $\delta_{Ni}$ represents the degradation rate of nitrogenase. We can effectively account for the lag introduced by intermediate processes not taken into account explicitly in the model by increasing the value of $\delta_\text{Ni}$ so that [Ni] relaxes more slowly. Assuming that nitrogenase produces fixed nitrogen at a constant rate,  we arrive at the equation that governs cN levels in cyanobacteria:
\begin{equation}
        \frac{d \text{[cN]}}{dt}=L'_n+v'_n \text{[Ni]} - \delta'_n \text{[cN]},
        \label{mo6}
\end{equation}
where $L'_n$ represents the flux of cN from the exterior of the cell. Assuming that the levels of cN relax rapidly we solve Equation \eqref{mo6} for the steady state. Substituting in Equation \eqref{mo5} we find:
\begin{equation}
\begin{aligned}
        \frac{d \text{[cN]}}{dt} =L_n+\frac{v_{n}^r \kappa_{n}^r \text{[HetR]}^2}{1+ \kappa_{n}^r \text{[HetR]}^2}-\delta_n \text{[cN]}\;,
\end{aligned}
        \label{mo9}
\end{equation}
where
\begin{equation}
        L_n=\frac{1}{\delta_n}\left(v_n L_\text{Ni}+\delta_\text{Ni}L'_n\right),\qquad v^r_n=\frac{v'_n}{\delta'_n}v^r_\text{Ni},\qquad\delta_n=\delta_\text{Ni},\qquad \kappa^r_n=\kappa^r_{Ni}.
        \label{mo10}
\end{equation}

\begin{table}[t]
{\centering
$\begin{array}{c|c|c|c}
        \multicolumn{4}{c}{\text{Constants}}\\\hline
l_a=0.2&l_r=0.01&
l_s=0.0001&l_n=0\\\hline
d_a=0.7&d_s=0.05&
d_n=0.01& \beta^a_a=4\\\hline
\beta_a^r=4&\beta_a^{ar}=8&
\beta_r^a=1&\beta_r^r=1\\\hline
\beta_r^{ar}=3&\beta_s^r=0.385&
\beta_n^r=0.06&\gamma_a^a=3\\\hline
\gamma_a^r=2.4&\gamma_s^r=1.2&
\multicolumn{2}{c}{\gamma_n^r=2.75}\\\hline
\end{array}$
\caption{Parameters for Eq.~\eqref{mo14} that reproduce heterocyst formation under noisy conditions and pattern formation when PatS and cN diffuse along a strain of cyanobacteria. \label{tmo1}}}
\end{table}

To get a closed system of equations, we shall investigate the relation between cN and 2-OG. Both are related by means of the GS/GOGAT cycle (Fig.~\ref{fig:gsgogat}). Assuming the cycle is in equilibrium and reactions are grounded on the law of mass action, the following two conditions must be satisfied:
\begin{equation}
        [\text{glutamate}]=\kappa_\leftarrow[\text{glutamine}][\text{2-OG}],\qquad [\text{glutamine}]=\kappa_\rightarrow[\text{glutamate}][\text{cN}],
        \label{mo11}
\end{equation}
that lead to the relation:
\begin{equation}
        \text{[2-OG]} =\frac{1}{\kappa_\leftarrow \kappa_\rightarrow\text{[cN]}}.
        \label{mo12}
\end{equation}
However, this expression does not behave properly for small concentrations of cN, which are expected under cN deprivation: 2-OG levels would increase without limit. In fact, 2-OG production is controlled by some processes that are not considered in this work and so its value must be limited. We can effectively include such a limiting value by means of a translation on [cN] in Eq.~\eqref{mo12}
\begin{equation}
        \text{[2-OG]} =\frac{1}{\kappa_\text{2-OG}+\kappa_\leftarrow \kappa_\rightarrow\text{[cN]}}\;,
        \label{mo13}
\end{equation}
which reaches the maximum value $\text{[2-OG]}_\text{max}=1/\kappa_\text{2-OG}$ at $\text{[cN]}=0$.

\medskip

Finally we can introduce the differential equations governing cyanobacterial reaction to nitrogen deprivation. They represent the temporal variation of the most important factors of the genetic circuit, namely NtcA, HetR, PatS and cN. Using the production rates \eqref{ntca}, \eqref{hetr}, \eqref{pats}, \eqref{mo9} and introducing degradation rates constants, $\delta_*$, we find:
\begin{equation}
        \begin{aligned}
                \frac{dq_a}{d\tau}&=l_a+\frac{\beta^a_a \gamma^a_a q^2_a+\beta_a^r \gamma_a^r q^2_r (1+q_{n})+\beta_a^{ar}\gamma_a^a q^2_a\gamma_a^r q^2_r}{(1+q_{n}+\gamma_a^a q^2_a)(1+\gamma_a^r q_r^2)}-d_a q_a,\\
                \frac{dq_r}{d\tau}&=l_r+\frac{\beta_r^a  q^2_a(1+q_s)+\beta_r^r  q^2_r(1+q_{n})+\beta_r^{ar}q^2_a q^2_r}{(1+q_{n}+q^2_a)(1+q_s+q_r^2)}- q_r,\\
                \frac{dq_s}{d\tau}&=l_s+\frac{\beta_s^r \gamma_s^r q^2_r}{1+\gamma_s^r q^2_r}-d_s q_s,\\
                \frac{dq_{n}}{d\tau}&=l_{n}+\frac{\beta^r_{n} \gamma^r_{n} q^2_r}{1+\gamma^r_{n} q^2_r}-d_{n} q_{n},\\
        \end{aligned}
        \label{mo14}
\end{equation}
where we have introduced the adimensional variables:
\begin{equation}
        q_a=\underbrace{\sqrt{\frac{\kappa_{r}^a}{\kappa_\text{2-OG}}}}_{\phi_a}\text{[NtcA]},\quad q_r=\underbrace{\sqrt{\kappa_r^r}}_{\phi_r}\text{[HetR]},\quad q_s = \underbrace{\kappa_r^s}_{\phi_s} \text{[PatS]},\quad q_{n} = \underbrace{\frac{\kappa_\leftarrow\kappa_\rightarrow}{\kappa_\text{2-OG}}}_{\phi_n} \text{[cN]},\qquad \tau = \delta_r t,\\
        \label{mo15}
\end{equation}
and the constants
\begin{equation}
l_* = \frac{L_* \phi_*}{\delta_r},\qquad \beta^{\text{\tiny $\bullet$}}_* = \frac{v^{\text{\tiny $\bullet$}}_* \phi_*}{\delta_r}, \qquad \gamma^{\text{\tiny $\bullet$}}_* = \frac{\kappa^{\text{\tiny $\bullet$}}_*}{\kappa^{\text{\tiny $\bullet$}}_r},\qquad d_* =  \frac{\delta_*}{\delta_r}.
\end{equation}

Let us finally stress that this is a \emph{deterministic} model for a single cyanobacterium. The study of the cyanobacterial strain is left to the final section. We advance that the main modification will be adding diffusion processes for the inhibitors PatS and cN through the chain.  An important ingredient in pattern formation, noise, will be also added to the equations.

\subsection*{Unicellular dynamics}

In this section we analyze the dynamical system \eqref{mo14} for a set of constants (Table \ref{tmo1}) that exhibit both the dynamical and the structural properties of heterocyst differentiation. Following the usual procedure in the theory of dynamical systems, we study the basic properties of equations \eqref{mo14}, such as fixed points and linear stability analysis, so to analyze the key features leading to heterocyst differentiation.

\begin{figure}[t]
\centering
\includegraphics[width=5.7in]{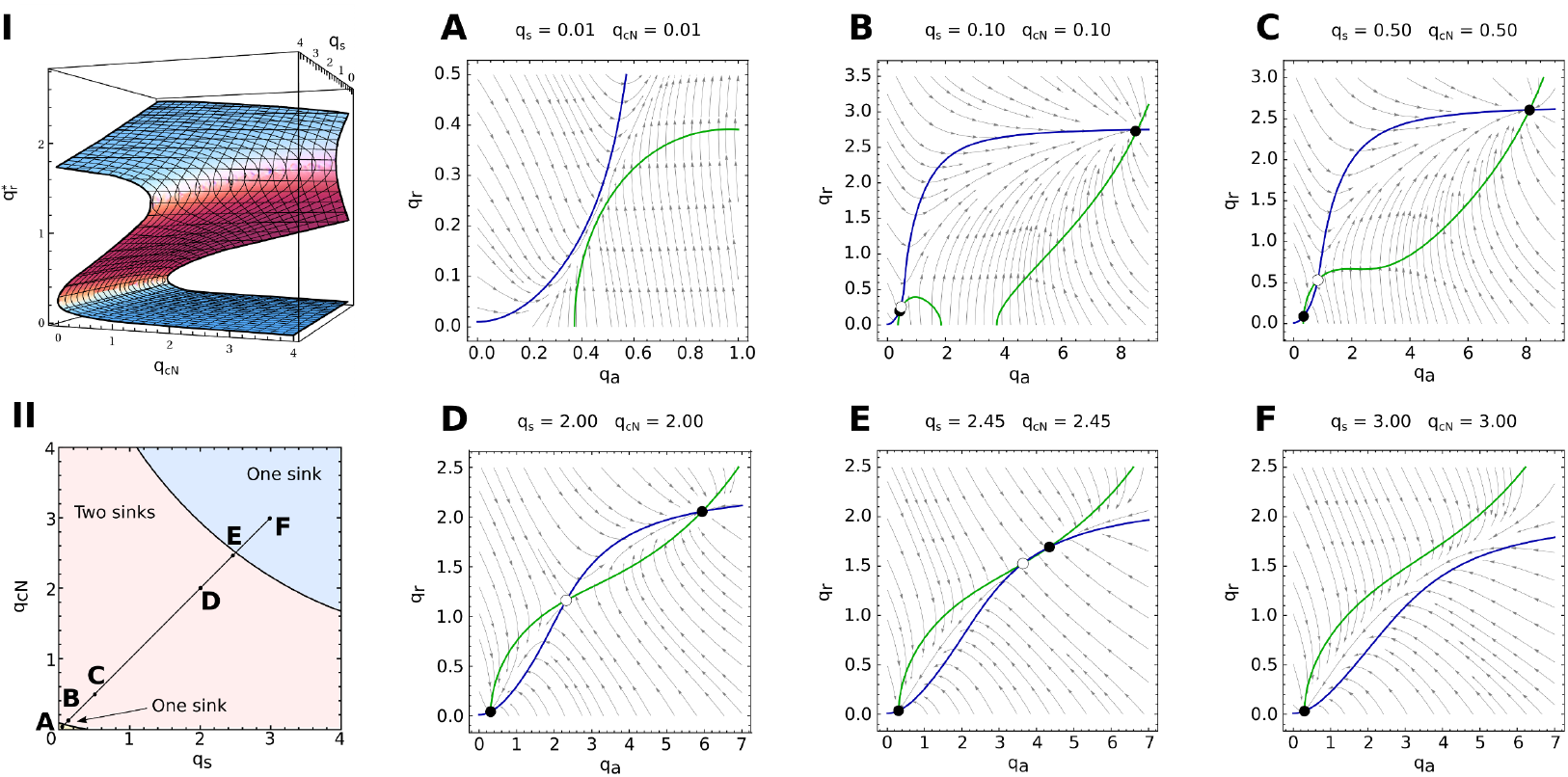}
\caption{Adiabatic elimination of the fast variables $q_r$ and $q_a$. Due to the fast dynamics that HetR and NtcA exhibit, we can approach the treatment of the system by adopting a point of view that follows the slower variables $q_s$ and $q_n$. From this viewpoint, the time-evolution of the pair $(q_s (t),q_n(t))$ is considered by assuming that $q_r$ and $q_a$ instantaneously relax to an equilibrium, which corresponds to a sink ($q_r^*$, $q_a^*$) for the fixed pair ($q_s(t)$, $q_n(t)$). Depending on the region of the $(q_s,q_n)$-plane, there are three fixed points (two sinks corresponding to the highest and the lowest concentrations respectively and a saddle in the middle) or one (a sink) for $q_r$ and $q_a$ (I and II). There are two one-sink regions that are separated from the two-sink region by saddle-node bifurcations (A-F). Sinks and saddles are represented by filled and unfilled circles respectively and arrows indicate the flow of the dynamics. We can then imagine the dynamics of $q_s$ and $q_n$ as evolving either in the bottom or in the top branch of I. In the two-sink region, both branches are plausible and the history of the dynamics determine the solution (hysteresis effect): a dynamics in a branch will continue in it until experiencing a bifurcation in the $(q_r,q_a)$ plane (see Fig.~\ref{fig:states} for examples).}
\label{fig:ad_el}
\end{figure}

Taking into account the difference between the relaxation times of the constituents of the model, given by the inverses of $d_*$ (see Table \ref{tmo1}), we can think it is composed of two temporally separated systems: a rapid one, formed by HetR and NtcA, showing fast dynamics that relaxes to its steady state almost instantaneously and a slow one, composed of PatS and cN, whose evolution is dictated by the values of HetR and NctA in their instantaneous equilibrium. This corresponds to an adiabatic elimination technique \cite{Manneville} that helps in understanding the behavior of the system since it reduces the complexity of the dynamical system by splitting it into two simpler interdependent subsystems.

First, we look for the fixed points of the fast variables $q_a$ and $q_r$ for each pair of values of $q_s$ and $q_n$
\begin{equation}
        f_a(q_s,q_n)=\frac{d q_a}{d\tau}=0, \qquad f_r(q_s,q_n)=\frac{d q_r}{d\tau}=0.
        \label{mo24}
\end{equation}
The numerical solution to this problem is sketched in Fig.~\ref{fig:ad_el}. We find three different branches of solutions that coexist in some regions. The fixed points on the lower and upper branches are always stable (blue region in Fig.~\ref{fig:ad_el}I) and those lying on the middle branch (red region) are saddles. Transitions between the regions with one and three fixed points correspond to saddle-node bifurcations in which the middle branch of solutions coalesce with the lower and the upper one respectively. The basins of attraction of both stable fixed points are separated by the stable manifold of the saddle point (Fig.~\ref{fig:ad_el} II).

In the bistable region the system behaves as a switch that can be either OFF in a vegetative-like state (lower branch, with a small production of HetR and NtcA) or ON in a heterocyst-like one (upper branch, with a high production of HetR and NtcA). A sufficiently large perturbation may result in the system crossing the manifold of the saddle and falling into the other stable branch of solutions. The distance between the saddle and the nodes determines the size of the perturbation needed to activate or inactivate the system.

\begin{figure}
\centering
\includegraphics[width=5.7in]{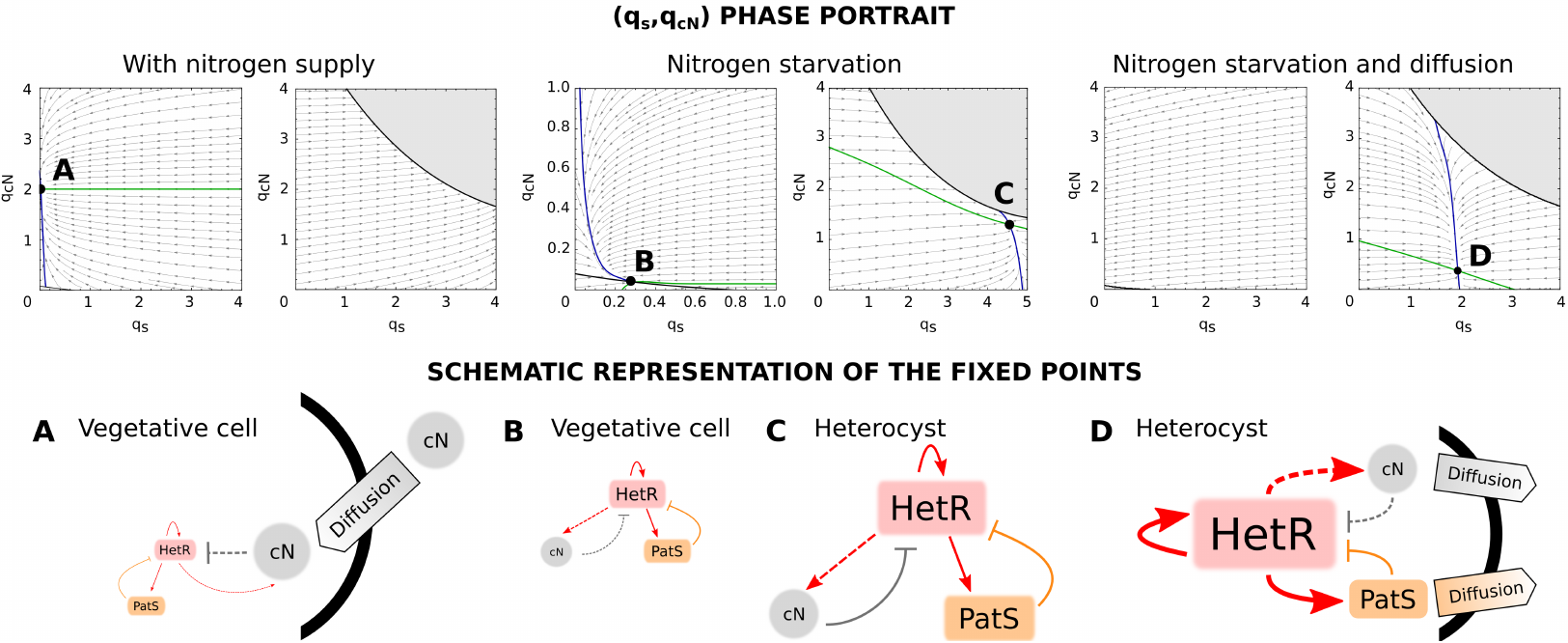}
\caption{States of a cyanobacterium when subjected to different conditions of nitrogen and diffusion. When the cell is provided of cN ($l_n = 0.03$), there is only one stable fixed point (A) in the bottom branch, which corresponds to a state in which the production of both HetR and PatS is minimum (vegetative-like state). When subjected to nitrogen deprivation ($l_n=0$), there are two stable fixed points (B and C) each one in a different branch. The first point (B) is a vegetative-like state in which there exists an equilibrium between a small production of HetR, PatS and cN. The same kind of equilibrium is present in the second fixed point (C) but in this case the production of all TFs and cN is high (heterocyst-like steady state). When the cell is exposed to nitrogen stress its trajectory evolves from A to the steady state B and thus it remains vegetative. Assuming some diffusion of cN and PatS from the cell ($l_s = -0.2$ and  $l_n=-0.002$), the only stable state (D) corresponds to a heterocyst-like state with high levels of production of HetR, cN and PatS, being the latter transported to the surroundings of the cell.}
\label{fig:states}
\end{figure}

Once \eqref{mo24} is solved, we can apply the solution to calculate the effective field sensed by the $(q_s,q_n)$ pair. In the regions showing bistability the field takes two very different forms, one corresponding to the values of the lower branch and another corresponding to those of the upper one (Fig.~\ref{fig:states}). We expect a {\it hysteresis effect}: if initially the dynamics lies on a particular branch it will remain on it unless a fluctuation or a bifurcation makes the system jump to the other branch. 

When supplied of cN (Fig.~\ref{fig:states}A) we find only one stable fixed point that corresponds to a vegetative-like state, since it belongs to the lower branch. The upper branch is completely unstable: all the dynamics lying on it will fall down to the lower branch and eventually be attracted to the vegetative-like sink. The steady state is very robust against perturbations since it is far from the bifurcation region and there is a significant distance to the saddle in the $q_r-q_a$ plane.

By reducing the flow of cN from the exterior of the cell ($l_n=0$) we find that a stable fixed point appears in the upper branch, a heterocyst-like state, and the vegetative-like state gets closer to the bifurcation region becoming more susceptible to perturbations that can make the system jump to the upper branch. When cN is eliminated from the media the cyanobacterium would evolve from state A to state B in the lower branch until a perturbation pushes it to the upper branch, eventually becoming an heterocyst due to the field acting on that branch (Fig.~\ref{fig:states}B and C). 

Diffusion protects cells in the neighborhood of the newly formed heterocyst to initiate the differentiation: as heterocysts are producers of cN and PatS, the vegetative fixed point of the cells in the neighborhood will move towards an A-like state, thus becoming more stable to perturbations. The heterocyst-like fixed point also becomes more stable due to diffusion, since its production of inhibitors is distributed among other cells (see Fig.~\ref{fig:states}D).

\subsection*{Strains of cyanobacteria. Heterocyst patterns}

In the previous section, we introduced a single cell model for the cyanobacterial reaction to nitrogen-limiting conditions. There we have shown that, for a specific range of parameters, the model exhibits features that would lead to heterocyst development under noisy conditions. Nevertheless, the model should be extended to cyanobacteria chains to account for heterocyst development since, as said above, isolated cyanobacteria do not become heterocysts by themselves; the action of the chain is needed to generate heterocysts.

In this section, we extend the previous results and consider a chain of vegetative cells facing nitrogen deprivation. The main modification is the introduction of diffusion of PatS and cN along the cyanobacteria chain. For this purpose we add to Equation \eqref{mo15} the discrete version of the diffusion equation:

\index{diffusion constant}
\begin{equation}
        \frac{d C_i}{dt}=D_C\left(C_{i+1}+C_{i-1}-2C_i\right).
        \label{st1}
\end{equation}
where $D_C$ is called the {\it diffusion constant} of the element C. Now, it is straightforward to introduce PatS and cN diffusion into the equations. The dynamics of cell $i$ is characterized by the following set of equations:
\begin{equation}
        \begin{aligned}
                \frac{dq_{i,a}}{d\tau}&=l_a+\frac{\beta^a_a \gamma^a_a q^2_a+\beta_a^r \gamma_a^r q^2_r (1+q_{n})+\beta_a^{ar}\gamma_a^a q^2_a\gamma_a^r q^2_r}{(1+q_{n}+\gamma_a^a q^2_a)(1+\gamma_a^r q_r^2)}-d_a q_{i,a} +G_{i,a}(t),\\
                \frac{dq_{i,r}}{d\tau}&=l_r+\frac{\beta_r^a  q^2_a(1+q_s)+\beta_r^r  q^2_r(1+q_{n})+\beta_r^{ar}q^2_a q^2_r}{(1+q_{n}+q^2_a)(1+q_s+q_r^2)}- q_{i,r}+G_{i,r}(t),\\
                \frac{dq_{i,s}}{d\tau}&=l_s+\frac{\beta_s^r \gamma_s^r q^2_r}{1+K_s^r q^2_r}-d_s q_{i,s}+D_s\left(q_{i+1,s}+q_{i-1,s}-2q_{i,s}\right)+G_{i,s}(t),\\
                \frac{dq_{i,n}}{d\tau}&=l_{n}+\frac{\beta^r_{n} \gamma^r_{n} q^2_r}{1+\gamma^r_{n} q^2_r}-d_{n} q_{i,n}+D_{n}\left(q_{i+1,n}+q_{i-1,n}-2q_{i,n}\right)+G_{i,n}(t),\\
        \end{aligned}
        \label{sty}
\end{equation}
which constitutes the model for a cyanobacteria strain. To account for environment variability we add white noise, $G_{i,*}(t)$, of the same amplitude, $\langle G_{i,*}(t) G_{i,*}(t')\rangle=\xi\delta(t-t')$, for all the components of the system. Based on these equations, we investigate the conditions that lead to a heterocyst pattern. It is easy to notice that they correspond to an activator-inhibitor system of cells coupled in a reaction-diffusion scheme \cite{morelli}. This kind of systems produce regular pattern formation \cite{turing, meinhardt}. Turing (linear stability) analysis of equations \eqref{sty} (see SI) provides insight on the periodicity of patterns. It is interesting to show that the minimum periodicity observed in such analysis is larger than 1, which means that a single bacteria is unable to differentiate.

\begin{figure}[t!]
\centering
\includegraphics[width=3.25in]{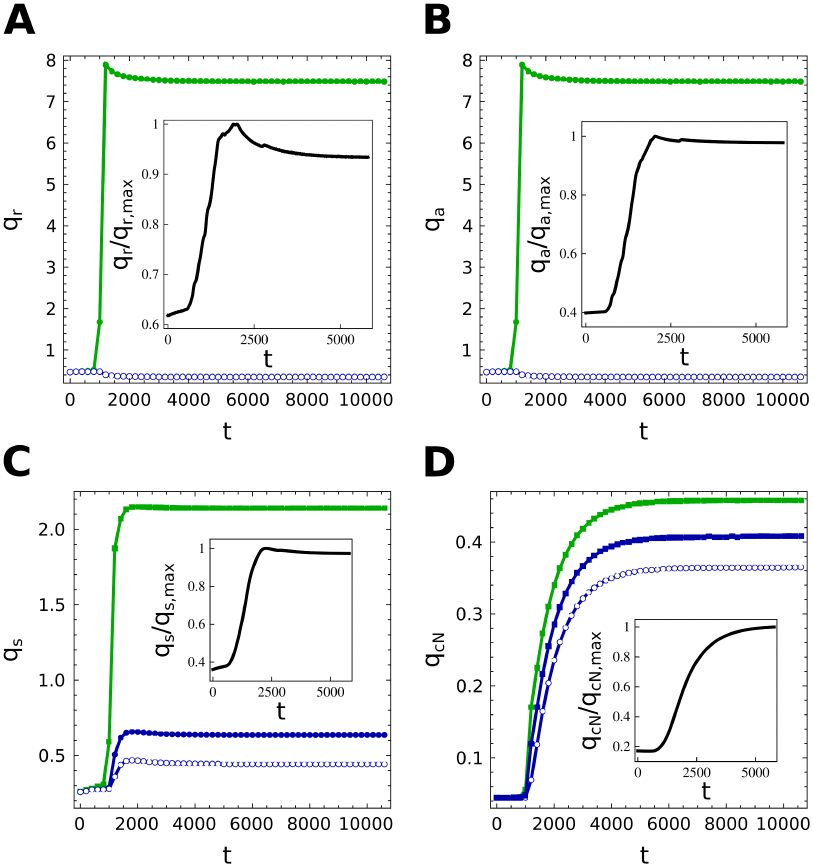}
\caption{Time evolution of the main components of the differentiation in heterocysts (green) and vegetative cells (blue). Averages along the strain are also presented (black). Heterocysts, due to the early diffusion, evolve toward steady states of the type D of Fig \ref{fig:states} characterized by high levels of HetR and NtcA while vegetative cells present very low concentrations of them (A and B). The levels of PatS and cN in vegetative cells depend on their distance to close heterocysts: C and D show the concentrations of PatS and cN in a heterocyst and in its first two neihbouring vegetative cells, which clearly highlight the effect of diffusion along the strain.}
\label{fig:time_ev} 
\end{figure}

\smallskip

We performed the direct integration of equations \eqref{sty} for chains of $200$ cyanobacteria. We used a Runge-Kutta method for the numerical integration of stochastic differential equations (see Methods) \cite{greenside}. The level of noise that best reproduces heterocyst pattern is $\xi=0.001$ for the set of parameters of Table \ref{tmo1}. Importantly, isolated cells do not initiate differentiation with this level of noise, in agreement with the results from the linear stability analysis. On the other hand, diffusion constants have been set to $D_s=0.1$ and $D_n=0.2$. Heterocysts patterns develop for different levels of noise and diffusion constants, but the model parameters, which characterize cell response to nitrogen deprivation, should change accordingly. This correlation between noise, diffusion and model parameters supports the idea that cyanobacteria have evolved towards the better response to the normal levels of noise in their environment.

In Fig.~\ref{fig:time_ev} we show the dynamics that the 4 variables exhibit when the chain is deprived of cN. We observe that the chain relaxes to the constant protein levels of the vegetative-like state we showed in the previous chapter. Then, due to the coupled action of noise and diffusion, some cells start to differentiate. As new forming heterocysts appear, their production and exportation of inhibitors to the surrounding cells make the latter more stable to perturbations stopping their differentiation. The model reproduces very well the initial peak that both NtcA and HetR present experimentally \cite{Huang2004,Jang2009}. PatS increases more slowly to its steady value reducing the levels of NtcA and HetR and, finally, cN is generated by heterocysts stabilizing the pattern.

\begin{figure}
\centering
\includegraphics[width=6.3in]{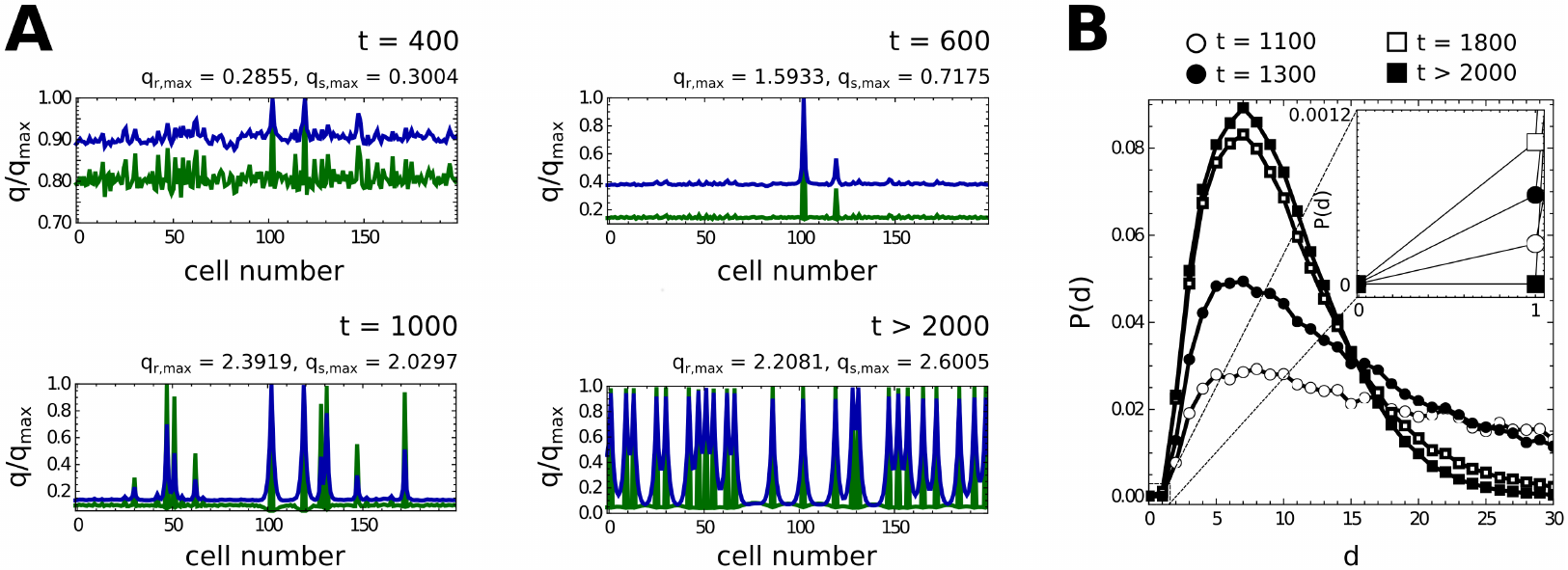}
\caption{Time-evolution of the pattern of heterocysts (A) and of the probability distribution of the distance between consecutive heterocysts (B). Green and blue curves represent the concentration profiles of HetR and PatS (NtcA and cN are not presented since their behavior along the strain is comparable to the one of HetR and PatS, see Fig.~\ref{fig:time_ev} to see the similarities). Small perturbations along the strain of vegetative cells (initially in the steady state B of Fig.~\ref{fig:states}) are amplified due to diffusion processes in a demonstration of Turing's theory \cite{turing}. New heterocysts appear in regions that are not dominated by the action of other heterocysts. Finally, the competition between nearby differentiating cells ceases the differentiation of some of them, as observed in B: consecutive heterocysts, which are created by strong perturbations, finally dissapear due to the aforementioned competition. The final pattern presents localized levels of HetR (heterocysts) and a diffusive-like behavior of PatS, as it was expected.}
\label{fig:strain}
\end{figure}

Fig.~\ref{fig:strain} shows the evolution of the profile for a 200 cells chain of cyanobacteria. We observe that heterocysts progressively appear in those regions in which others heterocysts do not have effect ({\em i.e.} those vegetative cells that are not supplied of sufficient cN and PatS). Finally, a semiregular pattern is generated. PatS and cN diffuse along the strain exhibiting smooth variations between vegetative cells and heterocysts, while HetR and NtcA present very abrupt variations between cell types.

Finally, in Fig.~\ref{fig:strain}.B we show the time-evolution of the histogram for the distance between two consecutive heterocysts. It should be stressed that although initially some close heterocysts appear, they are eliminated by the non-linear action of the system during the differentiation process. Close heterocysts compete for the same region of action (the same vegetative cells that consume their PatS and cN) and then they cannot reach the optimal heterocyst-like state, which is stable to slight perturbations. Finally, one of them falls down from the upper branch becoming a vegetative cell. This behavior is typically observed experimentally \cite{flores, Zhang2006}. The final histogram can be nicely fitted by a $\Gamma$-distribution.

\section*{Discussion}

The study of cell differentiation and its underlying mechanisms constitutes one of most intriguing problems in biology. This phenomena is the basis for multicellular organism and pattern formation. The approach we have chosen in this paper is to deal with a simple system, the heterocyst formation in cyanobacteria strains, yet complex enough to capture the main ingredients of some of the mechanism for cell differentiation and pattern formation under external driving. The knowledge of the basic regulatory genes and their corresponding interactions allows for a detailed description of cell dynamics. We have proposed the dynamical equations for these regulatory genes based in the statistical mechanics of the regulation process. This allow us to obtain a detailed description of the \emph{continuous time dynamics} of the main regulatory protein, in contrast to other discrete approximations based in boolean dynamics \cite{gerdtzen}. This kind of analysis have been proved successfully in other time dependent phenomena in the cyanobacterial word like circadian cycles \cite{teng}. The analysis of the "unicellular dynamics" has revealed that the two cellular stable states, vegetative and heterocyst cells, appear as attractors of the non-linear dynamics of the regulation equations. However, the study of many coupled cells is needed as cyanobacteria do not differenciate when isolated.

The model is thus completed by coupling a number of cells in a one-dimensional array so that combined nitrogen and PatS can diffuse along the chain of cells. One importaint ingredient affecting the dynamical behavior of the chain is noise, which appears to have a key role in the transition from the initial chain of vegetative cells to the steady state in which heterocysts coexist with vegetative cyanobacteria. Thus, the appearance of differentiation is, in our model, a pure stochastic event. The cooperative character of the strain is clear from the amount of noise needed to start the differenciation process which appears significantly smaller than that needed in isolated cells. The source of noise as well as its biological consequences is, nowadays a current topic of research \cite{suel}. In fact, at its initial state, differentiation of cells appear randomly along the strain, but shortly after its onset a characteristic distribution of heterocyst emerges. This distribution can be compared with the experimental one with a fairly agreement \cite{Yoon2001}. 

Although the model presented here integrates both internal cell dynamics and the coupling between cells via diffusion, there exist other ingredients that can also be incorporated. One issue that have not been considered in this work is the possibility of replication of vegetative cells. This effect has been taken into account in \cite{brown}. Although this improvement is relevant, it only affects, in our approach, to the mean separation between heterocysts, by opening a gap in the $\Gamma$-function shaped of Fig.~\ref{fig:time_ev}B and thus approaching better to the experimental distribution. 

Other improvements to the approach presented here will come from the availability of more experimental data. Unlike other approaches \cite{allard} where comparison is done (globally) with heterocyst distributions, our work would allow for a qualitative comparison of each component involved in the differentiation (see Fig.~\ref{fig:time_ev}). Unfortunately, experimental data are not enough to make a detailed fit and then extract reliable parameters. The availability of such data is extremely important both for having better set of model parameters and to validate new models. A complete understanding of the mechanism that derive in phenotypic differentiation is the first step for a modular comprehension of the whole cell \cite{Karr}.  


\section*{Methods}
To reproduce the dynamics of Eq.~\eqref{sty}  we make use of the integration scheme proposed in \cite{greenside}. Eq.~\eqref{sty} is a set of stochastic differential equations (SDE) so its numerical integration requires generating a statistical representative trajectory for a discrete set of time-values. A SDE of the form
\begin{equation} \label{SDE} \dot{x} = f(x) + G(t),\end{equation}
where $G(t)$ is a Gaussian white noise with
$$\langle G(t)\rangle = 0, \text{ and, } \langle G(t) G(t') \rangle = \xi \delta(t'-t),$$
can be integrated through a Runge-Kutta integration algorithm by adding a particular Gaussian signal at each stage of the scheme. This algorithm coincides with the usual Runge-Kutta scheme for $\xi=0$. In this work we have employed a $3_O4_S2_G$ algorithm, which is correct up to $3^{th}$ order, is developed in 4 stages and uses $2$ independent Gaussian random variables.

\section*{Acknowledgments}
We acknowledge financial support from the Spanish MINECO under projects FIS2011-25167 and FIS2012-38266-C02-01. J.G.G. is supported by the Spanish MINECO through the Ramon y Cajal program. We are grateful to S. Ares and J. Mu\~noz-Garc\'{\i}a for sharing ideas and useful discussions. We also acknowledge the "Genetic Regulation and Physiology of Cyanobacteria" group at University of Zaragoza for sharing insight on the genetic of heterocyst formation.  


\bibliography{ref}

\begin{thebibliography}{10}
\providecommand{\url}[1]{\texttt{#1}}
\providecommand{\urlprefix}{URL }
\expandafter\ifx\csname urlstyle\endcsname\relax
  \providecommand{\doi}[1]{doi:\discretionary{}{}{}#1}\else
  \providecommand{\doi}{doi:\discretionary{}{}{}\begingroup
  \urlstyle{rm}\Url}\fi
\providecommand{\bibAnnoteFile}[1]{%
  \IfFileExists{#1}{\begin{quotation}\noindent\textsc{Key:} #1\\
  \textsc{Annotation:}\ \input{#1}\end{quotation}}{}}
\providecommand{\bibAnnote}[2]{%
  \begin{quotation}\noindent\textsc{Key:} #1\\
  \textsc{Annotation:}\ #2\end{quotation}}
\providecommand{\eprint}[2][]{\url{#2}}

\bibitem{Koch1994}
Koch A, Meinhardt H (1994) {Biological pattern formation: from basic mechanisms
  to complex structures}.
\newblock Rev Mod Phys 66: 1481--1511.
\bibAnnoteFile{Koch1994}

\bibitem{Suel2006}
S\"{u}el GM, Garcia-Ojalvo J, Liberman LM, Elowitz MB (2006) {An excitable gene
  regulatory circuit induces transient cellular differentiation.}
\newblock Nature 440: 545--550.
\bibAnnoteFile{Suel2006}

\bibitem{wolk}
Wolk C (1998) {Heterocyst formation}.
\newblock Annu Rev Genet 30: 59--78.
\bibAnnoteFile{wolk}

\bibitem{flores}
Flores E, Herrero A (2010) {Compartmentalized function through cell
  differentiation in filamentous cyanobacteria}.
\newblock Nature Rev 8: 39--50.
\bibAnnoteFile{flores}

\bibitem{Schirrmeister2011}
Schirrmeister B, Antonelli A, Bagheri H (2011) {The origin of multicellularity
  in cyanobacteria}.
\newblock BMC Evol Biol 11: 45.
\bibAnnoteFile{Schirrmeister2011}

\bibitem{shi}
Shi Y, Zhao W, Zhang W, Ye Z, Zhao J (2006) {Regulation of intracellular free
  calcium concentration during heterocyst differenciation by HetR and NtcA in
  Anabaena sp. PCC7120}.
\newblock Proc Nat Acad Sci (USA) 103: 11334--11339.
\bibAnnoteFile{shi}

\bibitem{toepel}
Toepel J, Welsh E, Summerfield T, Pakrasi H, Sherman L (2008) {Differential
  Transcriptional Analysis of the Cyanobacterium Cyanothece sp. Strain {ATCC}
  51142 during Light-Dark and Continuous-Light Growth}.
\newblock J Bacteriol 190: 3904--3913.
\bibAnnoteFile{toepel}

\bibitem{Yoon1998}
Yoon H (1998) {Heterocyst Pattern Formation Controlled by a Diffusible
  Peptide}.
\newblock Science 282: 935--938.
\bibAnnoteFile{Yoon1998}

\bibitem{Zhang2006}
Zhang CC, Laurent S, Sakr S, Peng L, B\'{e}du S (2006) {Heterocyst
  differentiation and pattern formation in cyanobacteria: a chorus of signals.}
\newblock Mol microbiol 59: 367--375.
\bibAnnoteFile{Zhang2006}

\bibitem{kumar}
Kumar K, Mella-Herrera R, Golden J (2010) {Cyanobacterial heterocists}.
\newblock Cold Spring Harb Perspect Biol 2: a000315.
\bibAnnoteFile{kumar}

\bibitem{allard}
Allard J, Hill A, Rutenberg A (2007) {Heterocyst patterns without patterning
  proteins in cyanobacterial filaments}.
\newblock Dev Biol 312: 427--434.
\bibAnnoteFile{allard}

\bibitem{brown}
Brown A, Rutenberg A (2012) {Reconciling cyanobacterial fixed-nitrogen
  distributions and transport experiments with quantitative modelling}.
\newblock Phys Biol 9: 016007.
\bibAnnoteFile{brown}

\bibitem{gerdtzen}
Gedtzen Z, Salgado J, Osses A, Asenjo J, Rapaport I, et~al. (2009) {Modeling
  heterocyst pattern formation in cyanobacteria}.
\newblock BMC Bioinf 10: S16.
\bibAnnoteFile{gerdtzen}

\bibitem{Laurent2005}
Laurent S, Chen H, B\'{e}du S, Ziarelli F, Peng L, et~al. (2005)
  {Nonmetabolizable analogue of 2-oxoglutarate elicits heterocyst
  differentiation under repressive conditions in Anabaena sp. PCC 7120.}
\newblock P Natl Acad Sci USA 102: 9907--9912.
\bibAnnoteFile{Laurent2005}

\bibitem{Muro-Pastor2001}
Muro-Pastor MI, Reyes JC, Florencio FJ (2001) {Cyanobacteria perceive nitrogen
  status by sensing intracellular 2-oxoglutarate levels.}
\newblock J Biol Chem 276: 38320--38328.
\bibAnnoteFile{Muro-Pastor2001}

\bibitem{Vazquez-Bermudez2003}
V\'{a}zquez-Berm\'{u}dez MF, Herrero A, Flores E (2003) {Carbon supply and
  2-oxoglutarate effects on expression of nitrate reductase and
  nitrogen-regulated genes in Synechococcus sp. strain PCC 7942}.
\newblock FEMS Microbiol Lett 221: 155--159.
\bibAnnoteFile{Vazquez-Bermudez2003}

\bibitem{Muro-Pastor2005}
Muro-Pastor MI, Reyes JC, Florencio FJ (2005) {Ammonium assimilation in
  cyanobacteria.}
\newblock Photosynth Res 83: 135--150.
\bibAnnoteFile{Muro-Pastor2005}

\bibitem{Wei1994}
Wei TF, Ramasubramanian TS, Golden JW (1994) {Anabaena sp. strain PCC 7120 ntcA
  gene required for growth on nitrate and heterocyst development.}
\newblock J Bacteriol 176: 4473--4482.
\bibAnnoteFile{Wei1994}

\bibitem{Frias1994}
Fr\'{\i}as JE, Flores E, Herrero A (1994) {Requirement of the regulatory
  protein NtcA for the expression of nitrogen assimilation and heterocyst
  development genes in the cyanobacterium Anabaena sp. PCC 7120}.
\newblock Mol microbiol 14(4): 823--832.
\bibAnnoteFile{Frias1994}

\bibitem{Vazquez-Bermudez2002}
V\'{a}zquez-Berm\'{u}dez MF, Herrero A, Flores E (2002) {2-Oxoglutarate
  increases the binding affinity of the NtcA (nitrogen control) transcription
  factor for the Synechococcus glnA promoter.}
\newblock FEBS lett 512: 71--74.
\bibAnnoteFile{Vazquez-Bermudez2002}

\bibitem{Tanigawa2002}
Tanigawa R, Shirokane M, {Maeda Si} Si, Omata T, Tanaka K, et~al. (2002)
  {Transcriptional activation of NtcA-dependent promoters of Synechococcus sp.
  PCC 7942 by 2-oxoglutarate in vitro.}
\newblock P Natl Acad Sci USA 99: 4251--4255.
\bibAnnoteFile{Tanigawa2002}

\bibitem{Ramasubramanian1996}
Ramasubramanian TS, Wei TF, Oldham aK, Golden JW (1996) {Transcription of the
  Anabaena sp. strain PCC 7120 ntcA gene: multiple transcripts and NtcA
  binding.}
\newblock J Bacteriol 178: 922--926.
\bibAnnoteFile{Ramasubramanian1996}

\bibitem{Ramasubramanian1994}
Ramasubramanian TS, Wei TF, Golden JW (1994) {Two Anabaena sp. strain PCC 7120
  DNA-binding factors interact with vegetative cell- and heterocyst-specific
  genes.}
\newblock J Bacteriol 176: 1214--1223.
\bibAnnoteFile{Ramasubramanian1994}

\bibitem{Buikema1991}
Buikema WJ, Haselkorn R (1991) {Isolation and complementation of nitrogen
  fixation mutants of the cyanobacterium Anabaena sp. strain PCC 7120.}
\newblock J Bacteriol 173: 1879--1885.
\bibAnnoteFile{Buikema1991}

\bibitem{Buikema1991a}
Buikema WJ, Haselkorn R (1991) {Characterization of a gene controlling
  heterocyst differentiation in the cyanobacterium Anabaena 7120.}
\newblock Gene Dev 5: 321--330.
\bibAnnoteFile{Buikema1991a}

\bibitem{Ehira2006}
Ehira S, Ohmori M (2006) {NrrA directly regulates expression of hetR during
  heterocyst differentiation in the cyanobacterium Anabaena sp. strain PCC
  7120.}
\newblock J Bacteriol 188: 8520--8525.
\bibAnnoteFile{Ehira2006}

\bibitem{Wisen1999}
Wis\'{e}n S, Jiang F, Bergman B, Mannervik B (1999) {Expression and
  purification of the transcription factor NtcA from the cyanobacterium
  Anabaena PCC 7120.}
\newblock Protein expres purif 17: 351--357.
\bibAnnoteFile{Wisen1999}

\bibitem{Alfonso2001}
Alfonso M, Kirilovsky D (2001) {Redox Control of ntcA Gene Expression in
  Synechocystis sp . PCC 6803 . Nitrogen Availability and NtcA Protein 1}.
\newblock Plant Physiol 125: 969--981.
\bibAnnoteFile{Alfonso2001}

\bibitem{Li2003}
Li JH (2003) {An increase in the level of 2-oxoglutarate promotes heterocyst
  development in the cyanobacterium Anabaena sp. strain PCC 7120}.
\newblock Microbiology 149: 3257--3263.
\bibAnnoteFile{Li2003}

\bibitem{Buikema2001}
Buikema WJ, Haselkorn R (2001) {Expression of the Anabaena hetR gene from a
  copper-regulated promoter leads to heterocyst differentiation under
  repressing conditions.}
\newblock P Natl Acad Sci USA 98: 2729--2734.
\bibAnnoteFile{Buikema2001}

\bibitem{Khudyakov2004}
Khudyakov IY, Golden JW (2004) {Different functions of HetR, a master regulator
  of heterocyst differentiation in Anabaena sp. PCC 7120, can be separated by
  mutation.}
\newblock P Natl Acad Sci USA 101: 16040--16045.
\bibAnnoteFile{Khudyakov2004}

\bibitem{Zhou1998}
Zhou R, Wei X, Jiang N, Li H, Dong Y, et~al. (1998) {Evidence that HetR protein
  is an unusual serine-type protease.}
\newblock P Natl Acad Sci USA 95: 4959--4963.
\bibAnnoteFile{Zhou1998}

\bibitem{Huang2004}
Huang X, Dong Y, Zhao J (2004) {HetR homodimer is a DNA-binding protein
  required for heterocyst differentiation, and the DNA-binding activity is
  inhibited by PatS.}
\newblock P Natl Acad Sci USA 101: 4848--4853.
\bibAnnoteFile{Huang2004}

\bibitem{Muro-Pastor2002}
Muro-Pastor AM, Valladares A, Flores E, Herrero A (2002) {Mutual dependence of
  the expression of the cell differentiation regulatory protein HetR and the
  global nitrogen regulator NtcA during heterocyst development.}
\newblock Mol microbiol 44: 1377--1385.
\bibAnnoteFile{Muro-Pastor2002}

\bibitem{Valladares2004}
Valladares A, Muro-pastor AM, Herrero A, Flores E (2004) {The NtcA-Dependent P
  1 Promoter Is Utilized for glnA Expression in N 2 -Fixing Heterocysts of
  Anabaena sp . Strain PCC 7120}.
\newblock J Bacteriol 186: 7337--7343.
\bibAnnoteFile{Valladares2004}

\bibitem{Fiedler2001}
Fiedler G, Muro-Pastor A (2001) {NtcA-Dependent Expression of the devBCAOperon,
  Encoding a Heterocyst-Specific ATP-Binding Cassette Transporter in Anabaena
  spp.}
\newblock J Bacteriol 183: 3795--3799.
\bibAnnoteFile{Fiedler2001}

\bibitem{Hebbar2000}
Hebbar PB, Curtis SE (2000) {Characterization of devH, a gene encoding a
  putative DNA binding protein required for heterocyst function in Anabaena sp.
  strain PCC 7120.}
\newblock J Bacteriol 182: 3572--3581.
\bibAnnoteFile{Hebbar2000}

\bibitem{Wu2004}
Wu X, Liu D, Lee MH, James W, Golden JW (2004) {patS Minigenes Inhibit
  Heterocyst Development of Anabaena sp . Strain PCC 7120 patS Minigenes
  Inhibit Heterocyst Development of Anabaena sp . Strain PCC 7120}.
\newblock J Bacteriol 186: 6422--6429.
\bibAnnoteFile{Wu2004}

\bibitem{Yoon2001}
Yoon Hs, Golden JW (2001) {PatS and Products of Nitrogen Fixation Control
  Heterocyst Pattern PatS and Products of Nitrogen Fixation Control Heterocyst
  Pattern}.
\newblock J Bacteriol 183: 2605--2613.
\bibAnnoteFile{Yoon2001}

\bibitem{Fay1992}
Fay P (1992) {Oxygen relations of nitrogen fixation in cyanobacteria.}
\newblock Microbiol Rev 56: 340--373.
\bibAnnoteFile{Fay1992}

\bibitem{Wolk1974}
Wolk CP, Austin SAMM, Bortins J, Galonsky A (1974) {Autoradiographic
  localization of {$^{13}$}N after fixation of {$^{13}$}N-labeled nitrogen gas
  by a heterocyst-forming blue-gree alga}.
\newblock J Cell Biol 61: 440--453.
\bibAnnoteFile{Wolk1974}

\bibitem{hwa}
Buchler NE, Gerland U, Hwa T (2003) {On schemes of combinatorial transcription
  logic}.
\newblock Proc Natl Acad Sci 100: 5136- 5141.
\bibAnnoteFile{hwa}

\bibitem{Bintu2005}
Bintu L, Buchler N, Garcia H, Gerland U, Hwa T, et~al. ({2005})
  {Transcriptional regulation by the numbers: models}.
\newblock {Current opinion in genetics \& development} {15}: {116-124}.
\bibAnnoteFile{Bintu2005}

\bibitem{Phillips2012}
Phillips R, Kondev J, Theriot J, Garcia H (2012) Physical Biology of the Cell.
\newblock New York: Garland Science, second edition.
\bibAnnoteFile{Phillips2012}

\bibitem{Manneville}
Manneville P (1990) Dissipative structures and weak turbulence.
\newblock London: Academic Press.
\bibAnnoteFile{Manneville}

\bibitem{morelli}
Morelli LG, Uriu K, Ares S, Oates AC (2013) {Computational Approaches to
  Developmental Patterning}.
\newblock Science 336: 187-191.
\bibAnnoteFile{morelli}

\bibitem{turing}
Turing AM (1952) {The Chemical Basis of Morphogenesis}.
\newblock Philos Trans R Soc London Ser B 237: 37-62.
\bibAnnoteFile{turing}

\bibitem{meinhardt}
Meinhardt H (1982) Models of Biological Pattern Formation.
\newblock London: Academic Press.
\bibAnnoteFile{meinhardt}

\bibitem{greenside}
{Greenside, HS and Helfand, E} ({1981}) {Numerical-Integration of Stochastic
  Differential Equations 2.}
\newblock {Bell System Technical Journal} {60}: {1927-1940}.
\bibAnnoteFile{greenside}

\bibitem{Jang2009}
Jang J, Shi L, Tan H, Janicki A, Zhang CC ({2009}) {Mutual Regulation of ntcA
  and hetR during Heterocyst Differentiation Requires Two Similar PP2C-Type
  Protein Phosphatases, PrpJ1 and PrpJ2, in Anabaena sp Strain PCC 7120}.
\newblock {J Bacteriol} {191}: {6059-6066}.
\bibAnnoteFile{Jang2009}

\bibitem{teng}
Teng SW, Mukherji S, Moffitt JR, de~Buyl S, O’Shea EK (2013) Robust circadian
  oscillations in growing cyanobacteria require transcriptional feedback.
\newblock Science 340: 737-740.
\bibAnnoteFile{teng}

\bibitem{suel}
Suel GM, Kulkarni RP, Dworkin J, Garcia-Ojalvo J, Elowitz MB (2007) Tunability
  and noise dependence in differentiation dynamics.
\newblock Science 315: 1716-1719.
\bibAnnoteFile{suel}

\bibitem{Karr}
Karr J, Sanghvi J, Macklin D, Gutschow M, Jacobs J, et~al. (2012) A whole-cell
  computational model predicts phenotype from genotype.
\newblock Cell 150: 389 - 401.
\bibAnnoteFile{Karr}

\end{thebibliography}

\section*{Supporting Information}

\subsection*{Regulatory equations: a statistical mechanics approach}

To derive the equations regulating transcription processes during heterocyst differentiation, we follow an approach similar to the detailed in \cite{hwa, Bintu2005, Phillips2012}. This is a thermodynamic approach to transcription in which binding sites are considered two-states systems, either empty or containing a binding protein. The probability that a given transcription factor (TF) is bound to its binding site is given by the Arrhenius formula
\begin{equation}
\label{arrhenius}
p_\text{TF} = \frac{\text{[TF]}K_\text{TF}}{1+\text{[TF]}K_\text{TF}}=\frac{q_\text{TF}}{1+q_\text{TF}}
\end{equation} 
where [TF] is the concentration of the TF, $K_\text{TF}$ is the inverse of the effective dissociation constant, which represents the concentration of half-maximal occupation, and $q_\text{TF} = \text{[TF]}K_\text{TF}$ is called the binding affinity. The denominator of Eq.~\eqref{arrhenius} is nothing but the canonical partition function of the promoter $Z=1+\text{[TF]}K_\text{TF}$, representing the Boltzmann-weighted sum of possible states of the binding site. Transcription starts with the binding of RNAp, which in the absence of interactions with TFs follows the probability law of Eq.~\eqref{arrhenius}:
\begin{equation}
p_\text{RNAp} =\frac{q_\text{RNAp}}{1+q_\text{RNAp}}
\end{equation}
Let us examine the case in which RNAp interacts with a TF within the promoter. In this case the partition function is
\begin{equation}
Z_\text{RNAp,TF} = 1 + \text{[TF]}K_\text{TF} + \text{[RNAp]}K_\text{RNAp} + \text{[RNAp]} \text{[TF]}K_\text{RNAp,TF} 
\end{equation}
with $K_\text{RNAp,TF}$ the inverse dissociation constant of the complex RNAp\&TF that can be higher than $K_\text{TF}K_\text{RNAp} $ if the interaction of the two proteins within the promoter is attractive or smaller if the interaction is repulsive. In the first case we say that the TF is an inhibitor while in the second case we say that the TF is an activator. We are interested in the probability that RNAp is bound to its binding site. This probability is
\begin{equation}
\label{arrhenius2}
\begin{aligned}
p_\text{RNAp} &= \frac{ \text{[RNAp]}K_\text{RNAp} + \text{[RNAp]} \text{[TF]}K_\text{RNAp,TF} }{1 + \text{[TF]}K_\text{TF} + \text{[RNAp]}K_\text{RNAp} + \text{[RNAp]} \text{[TF]}K_\text{RNAp,TF} } =\\
&=\frac{\displaystyle
	q_\text{RNAp} \left(1+q_\text{TF}\omega_\text{RNAp,TF}\right)}{\displaystyle1+q_\text{TF}+	q_\text{RNAp}\left(1+q_\text{TF}\omega_\text{RNAp,TF}\right)},
\end{aligned}
\end{equation} 
where we have used the definitions $q_\text{TF} = \text{[TF]}K_\text{TF}$, $q_\text{RNAp} = \text{[RNAp]}K_\text{RNAp}$ and $\omega_\text{RNAp,TF} = K_\text{RNAp,TF}/(K_\text{TF}K_\text{RNAp})$. Now, assuming that [RNAp] does not vary during the transcription process (or that it is not the limiting factor of transcription) and that transcription (which we remind is nothing but the production of mRNA) takes place at a given velocity $\nu$ when RNAp is bound to the promoter, we can find the relation between $q_\text{TF}$ and the transcription velocity $v_\text{mRNA}$:
\begin{equation}
\label{activator}
v_\text{mRNA} = L^\text{A} +v_{\text{TF}} \frac{q_\text{TF} \kappa^\text{A}_\text{TF}}{1+q_\text{TF}\kappa^\text{A}_\text{TF}},
\end{equation}
for the case in which the TF is an activator ($\omega_\text{RNAp,TF}>1$) or
\begin{equation}
v _\text{mRNA} = L^\text{I}  +v_{\text{TF}} \frac{1}{1+q_\text{TF}\kappa^\text{I}},
\end{equation}
in the case of an inhibitor ($\omega_\text{RNAp,TF}<1$). The remaining constants are
\begin{equation}
\begin{aligned}
v_{\text{TF}}&=\nu \frac{q_\text{RNAp}\left|\omega_\text{RNAp,TF}-1\right|}{(1+q_\text{RNAp})(1+q_\text{RNAp}\omega_\text{RNAp,TF})},\\
\kappa^\text{A}=\frac{1}{\kappa^\text{I}}&=\frac{1+\omega_\text{RNAp,TF} q_\text{RNAp}}{1+q_\text{RNAp}},\\
L^\text{A}_\text{mRNA}&=\nu\frac{q_\text{RNAp}}{1+q_\text{RNAp}},\\
 L^\text{I}_\text{mRNA}&=\nu\frac{q_\text{RNAp} \omega_\text{RNAp,TF}}{1+q_\text{RNAp}\omega_\text{RNAp,TF}}.
\end{aligned}
\end{equation}
These equations are of the form of the Michaelis-Menten equations of reaction kinetics with a leak term, represented by $L^A$ or $L^I$ depending on the case, that stands for the production of mRNA in the absence of regulation. This example shows the main features of the statistical mechanics approach to transcription. More complex transcription processes can be dealt with in a similar way by computing their corresponding partition function and counting the RNAp-active states. 

For all processes in the article we assume that TFs do not interact within the promoter but rather that they cooperatively affect the velocity at which transcription takes place. This is the simplest way in which we can consider the interaction between different TFs and, on the other hand, it is rich enough to represent the main features of the transcription processes we need to account for.

Let us sketch, for instance, the regulation of {\it ntcA} in heterocyst development (see main text). {\it ntcA} is regulated partly by NtcA (in its dimer configuration) and 2-OG and also by HetR. The partition function in this case is:
\begin{equation}
\begin{aligned}
Z_\text{RNAp,NtcA\&2-OG,HetR} &=1 + \text{[RNAp]}K_\text{RNAp} + \text{[NtcA]}^2\text{[2-OG]}K_\text{NtcA2,2-OG}+ \\
&+ \text{[RNAp]} \text{[NtcA]}^2\text{[2-OG]}K_\text{RNAp,NtcA2,2-OG} \\
&+ \text{[HetR]}^2K_\text{HetR2} + \text{[RNAp]} \text{[HetR]}^2K_\text{RNAp,HetR2} +\\
& + \text{[HetR]}^2 \text{[NtcA]}^2\text{[2-OG]}K_\text{NtcA2,2-OG}K_\text{HetR2}\\
& + \text{[RNAp]} \text{[HetR]}^2 \text{[NtcA]}^2\text{[2-OG]}K_\text{RNAp,NtcA2,2-OG}K_\text{RNAp,HetR2}
 \end{aligned}
\end{equation}
Both NtcA and HetR acts as activators, and we finally arrive to the transcription velocity
\begin{equation}
v_a=L_a+\frac{v^a_a \kappa^a_a [\text{2-OG}][\text{NtcA}]^2+v_a^r \kappa_a^r [\text{HetR}]^2+v_a^{ar} \kappa_a^a\kappa_a^r[\text{2-OG}][\text{NtcA}]^2[\text{HetR}]^2}{ (1+\kappa_a^a [\text{2-OG}][\text{NtcA}]^2)(1+\kappa_a^r [\text{HetR}]^2)}
\label{ntca}
\end{equation}
where $v_a^a$, $v_a^r$ and $v_a^{ar}$ represent the effective transcription velocity when NtcA, HetR or both are bound to DNA respectively. The constants $\kappa$ are obtained from $K$ by eliminating $q_\text{RNAp}$ from the equations following a procedure similar to that of Eq.~\eqref{activator} and can be thought as the inverse of effective dissociation constants associated with the binding of the different compounds.

Finally, we consider translation (the process by which mRNA is transformed into the corresponding protein) is produced at a constant rate $\eta$ per mole of mRNA. Therefore, the concentration of a given TF is given by:
\begin{equation}
\frac{d [\text{TF}]}{dt} = \eta_\text{TF} [\text{mRNA}_{\text{TF}}] -\delta_{\text{TF}} [{\text{TF}}],
\end{equation}
with $\delta_{\text{TF}}$ the inverse of the mean lifetime of the TF. On the other hand, the dynamics for concentration of mRNA is
\begin{equation}
\frac{d [\text{mRNA}_\text{TF}]}{dt} = v_{\text{mRNA}_\text{TF}} -\delta_{\text{mRNA}_\text{TF}} [\text{mRNA}_{\text{TF}}],
\end{equation}
with $\delta_{\text{mRNA}_\text{TF}}$ the inverse of the mean lifetime of the mRNA. Usually, $\delta_{\text{mRNA}_\text{TF}}>>\delta_{\text{TF}}$ and, as we are interested in [TF], we can assume that from the viewpoint of the characteristic dynamics of the TF, [mRNA] relaxes instantaneously to its equilibrium value:
\begin{equation}
[\text{mRNA}_{\text{TF}}]=\frac{v_{\text{mRNA}_\text{TF}}}{\delta_{\text{mRNA}_\text{TF}}},
\end{equation}
so
\begin{equation}
\label{eff}
\frac{d [\text{TF}]}{dt} = \eta \frac{v_{\text{mRNA}_\text{TF}}}{\delta_{\text{mRNA}_\text{TF}}} -\delta_{\text{TF}} [{\text{TF}}].
\end{equation}
In an abuse of notation, we redefine the variables $v_*^{\text{\tiny $\bullet$}}$ and $l_*$ as $v_*^{\text{\tiny $\bullet$}} \eta_*/\delta_{\text{mRNA}_*}$ and $v_*^{\text{\tiny $\bullet$}} \eta_*/\delta_{\text{mRNA}_*}$ respectively to express the effective constants in Eq.~\eqref{eff}. For instance, we find
\begin{equation} 
\frac{d[\text{NtcA}]}{dt} = L_a + \frac{v^a_a \kappa^a_a [\text{2-OG}][\text{NtcA}]^2+v_a^r \kappa_a^r [\text{HetR}]^2+v_a^{ar} \kappa_a^a\kappa_a^r[\text{2-OG}][\text{NtcA}]^2[\text{HetR}]^2}{ (1+\kappa_a^a [\text{2-OG}][\text{NtcA}]^2)(1+\kappa_a^r [\text{HetR}]^2)} - \delta_a [\text{NtcA}]
\end{equation}

\subsection*{Turing linear stability analysis}

Here we discuss the Turing stability analysis of our system \cite{Turing, Murray}. Let us first discuss the general theory. We consider a chain composed of $n$-dimensional dynamical systems, each one $i=1,\dots,n$ characterized by $\dot{\bm{q}}_i=\boldsymbol{f}(\boldsymbol{q}_i)$. If we add diffusion along the chain, the behavior of cell $i$ is characterized by:
\begin{equation}
\frac{d \boldsymbol{q}_i}{d\tau}= \boldsymbol{f}(\boldsymbol{q}_i) + \widetilde{D} (\boldsymbol{q}_{i+1}+\boldsymbol{q}_{i-1}-2\boldsymbol{q}_i),
\label{st3}
\end{equation}
where $\widetilde{D}$ is the {\it diffusion tensor}. Now we consider a (stable) fixed point, $\boldsymbol{q}_0$, of the dynamical system, i.e. $\boldsymbol{f}(\boldsymbol{q}_0)=\boldsymbol{0}$. It also constitutes a fixed point for the entire chain since diffusion terms cancel ($\boldsymbol{q}_i=\boldsymbol{q}_j$ for all $i$ and $j$). We want to analyze the effect of a small perturbation, $\boldsymbol{\Delta}$, around the steady state of the chain. Introducing the variables
\index{diffusion tensor}
\begin{equation}
\boldsymbol{q}_i= \boldsymbol{q}_0+\boldsymbol{\Delta}_i,
\label{st4}
\end{equation}
into Equation \ref{st3}, and expanding up to first order in $\boldsymbol{\Delta}$ one gets:
\begin{equation}
\frac{d \boldsymbol{\Delta}_i}{d\tau}= \nabla \boldsymbol{f}(\boldsymbol{q_0}) \cdot \boldsymbol{\Delta}_i+ \widetilde{D} (\boldsymbol{\Delta}_{i+1}+\boldsymbol{\Delta}_{i-1}-2\boldsymbol{\Delta}_i),
\label{st5}
\end{equation}
where $\nabla \boldsymbol{f}(\boldsymbol{q_0})$ is the Jacobian matrix of the field evaluated in the point $\boldsymbol{q}_0$. Furthermore, we can decompose the perturbation in terms of plane waves:
\begin{figure}[t]
	{\centering
		\includegraphics[width=0.4075\linewidth]{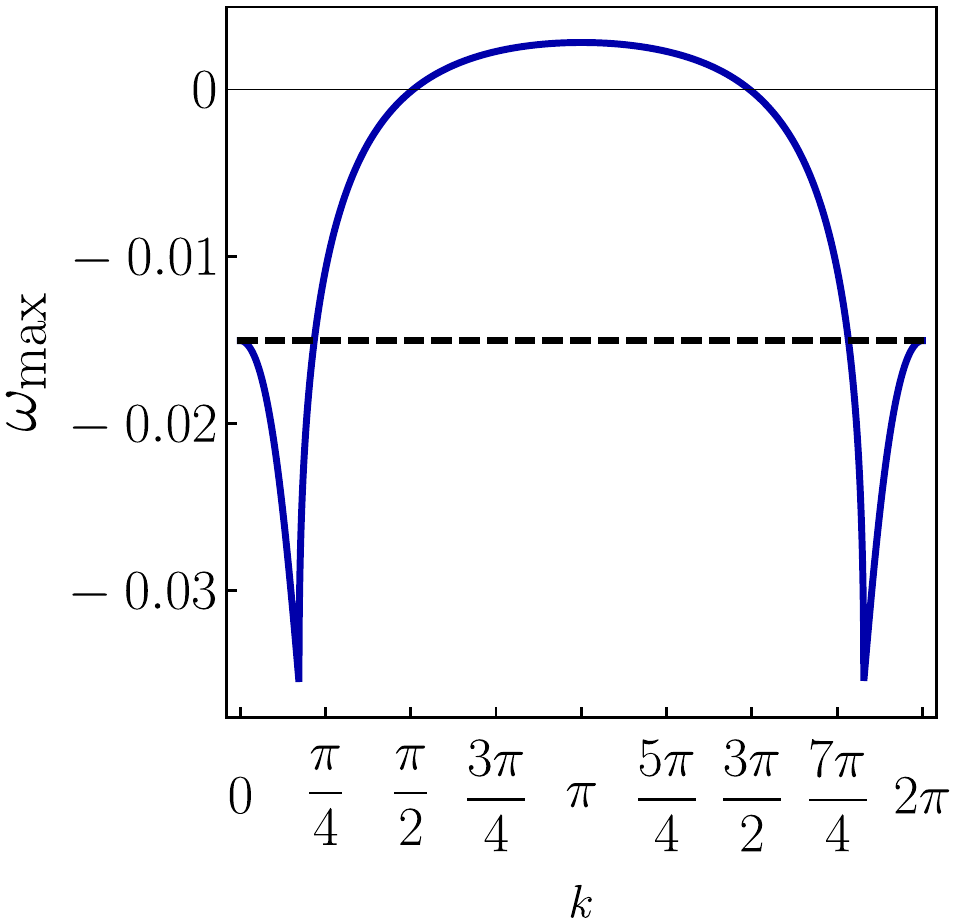}
		\caption{\label{figchain1} Results for $\omega_{\max}$ extracted from Equation \ref{st8} for the cyanobacterial system with  $D_s=0.1$ and $D_n=0.2$ (Blue). The black dashed line indicates the value at equilibrium in the absence of diffusion}}
\end{figure}
\begin{equation}
\begin{aligned}
\boldsymbol{\Delta}_i(\tau)&= \sum_k \boldsymbol{\Delta}_{i,k},\\
\boldsymbol{\Delta}_{i,k}(\tau)&= \boldsymbol{A} e^{\omega_k\tau} \cos (ki).
\end{aligned}
\label{st6}
\end{equation}
The admissible values of the wavevector $k$ depend on the length of the chain and on the boundary conditions. For instance, $k=n\pi/L$ with $n=1,2,3,\hdots,L$ for Von Neumann (zero flux) boundary conditions.  Introducing \ref{st6} into \ref{st5} we find:
\begin{equation}
\omega_k \boldsymbol{A}= \nabla\boldsymbol{f}(\boldsymbol{q_0})\boldsymbol{A}+2\widetilde{D}\boldsymbol{A}(\cos k-1),
\label{st7}
\end{equation}
which has nontrivial solutions if
\begin{equation}
\det(\nabla\boldsymbol{f}(\boldsymbol{q_0})+2\widetilde{D}(\cos k-1)-\omega_k)=0.
\label{st8}
\end{equation}
Therefore, the $k$ mode is related to some possible frequencies $\omega_k$. If those frequencies satisfy $\text{Re}(\omega_k)<0$, it is expected that the perturbation $\bm{\Delta}$ will be damped (negative exponential) and the system will recover its initial equilibrium. Nevertheless, if $\text{Re}(\omega_k)>0$ perturbations will be amplified (positive exponential) and thus it is expected that structures of wavelength $2\pi/k$ will develop. The important value that determines if the system is stable to perturbations of some wavevector $k$ is the largest real part of the $\omega_k$ from \ref{st8}.

We can now turn to our cyanobacteria chain characterized by Eq.~(16) of the main text. We analyze the effect of perturbations around the vegetative-like state for $D_s=0.1$ and $D_{n}=0.2$ (Figure \ref{figchain1}). We find that the system is unstable against perturbations of intermediate wavevectors, with an upper bound (representing a minimum length at which patterns can be formed) and a lower bound (maximum length). The value of the minimum length (measured in cell number) $l_\text{min} \approx 8/7 > 1$ so that a single cell is unable to differentiate.

\end{document}